\documentclass[12pt,preprint]{aastex}
\usepackage[twocolumn]{emulateapj5}

\shorttitle{VLT Observations of NGC 1313 X-2}
\shortauthors{Mucciarelli et al.}

\begin{document}

\title{VLT Observations of the Ultraluminous X-ray Source NGC 1313 X-2}

\author{P. Mucciarelli\altaffilmark{1,2}, L. Zampieri\altaffilmark{2},
R. Falomo\altaffilmark{2}, R. Turolla\altaffilmark{3} and A.
Treves\altaffilmark{4}}
\altaffiltext{1}{Department of Astronomy, University of Padua, Vicolo 
dell'Osservatorio 2, I-35122 Padova,
Italy; mucciarelli@pd.astro.it}
\altaffiltext{2}{INAF-Astronomical Observatory of Padua, Vicolo dell'Osservatorio 5,
I-35122 Padova, Italy; zampieri@pd.astro.it, falomo@pd.astro.it}
\altaffiltext{3}{Department of Physics, University of Padua, Via Marzolo 8, I-35131
Padova, Italy; turolla@pd.infn.it}
\altaffiltext{4}{Department of Physics and Mathematics, University of Insubria,
Via Valleggio 11, I-22100 Como, Italy; treves@mib.infn.it}

\begin{abstract}
We present archive ESO VLT photometric and spectroscopic data of the
Ultraluminous X-ray source NGC 1313 X-2. The superb quality of the VLT
images reveals that two distinct objects, with $R$ magnitudes 23.7 and
23.6, are visible inside the {\em Chandra} error box. The two objects,
separated by $0.75\arcsec$, were unresolved in our previous ESO 3.6
m+EFOSC image. We show that both are stars in NGC 1313, the first a
B0-O9 main sequence star of $\sim 20 M_\odot$, while the second a G
supergiant of $\sim 10 M_\odot$. Irrespectively of which of the two
objects the actual counterpart is, this implies 
that NGC 1313 X-2 is a high mass X-ray binary with a very massive donor.
\end{abstract}


\section{Introduction}

First revealed by {\em Einstein}, the population of ultraluminous
X-ray Sources (ULXs) has increasingly grown up in the last decade
mainly thanks to the observations of {\em ROSAT}
(e.g. \citealt{col02}), {\em XMM-Newton} (e.g. \citealt{fosc02a}) and
{\em Chandra} (e.g. \citealt{liu05,swa04}). About 150 ULXs are
included in the recent {\em Chandra} catalogue of \cite{swa04}. These
point-like sources have X-ray luminosities $L_{X} \gtrsim 10^{39}$
erg~s$^{-1}$, in excess of that of a $\sim 10 M_{\odot}$ compact
object accreting at the Eddington limit.  Variability in the X-ray
flux on timescales of months is observed in about half of the {\em
ROSAT} ULXs with multiple observations \citep{col02}, while about
5-15\% of the {\em Chandra} ULXs show variability during a single
observation (average exposure time $\sim$ 40 ks, \citealt{swa04}).

For several sources with sufficiently good statistics, the best fit to
the X-ray spectrum is obtained with a two-component model, a soft
multicolor disk (MCD) blackbody plus a power law. Some ULXs show
typical temperatures of the MCD component 5-10 times lower than those
of Galactic X-ray binaries. The high luminosity, the very soft thermal 
component (if it represents
the emission from a cool accretion disk) and the variability suggest
that these sources may be powered by accretion onto an Intermediate
Mass Black Hole (IMBH) of 100-1000 $M_{\odot}$. Nevertheless, many of
the ULX properties can be explained if they do not emit isotropically
\citep{kin01} or are dominated by emission from a relativistic jet
(e.g. \citealt{kaa03}). In this case, they may harbor stellar mass BHs
and may be similar to Galactic black hole binaries.

Multiwavelength observations are definitely a powerful tool to investigate
the nature of ULXs. Radio emission, when present, gives important clues on the
geometry, energetics and lifetime of ULXs \citep{kaa03,miln05}.
Optical follow-ups are crucial to identify ULX counterparts and clean up
the population from the significant contamination of background AGNs
and interacting SNe \citep{fosc02,mas03,swa04}. Up to now only a very
small number of ULXs have been convincingly associated with stellar objects
of known spectral type \citep[e.g.][]{liu02,liu04,kaa04}.
All these ULXs are hosted in star-forming regions and their optical counterparts have
properties consistent with those of early type O-B stars. Some ULXs are
also associated with extended optical emission nebulae \citep{pak02}.

NGC 1313 X-2 is a prototypical ULX \citep[see][and references therein]{mil03,zamp04,tur05}. With a luminosity $L_X \sim 10^{40}$erg
s$^{-1}$ in the 0.2-10.0 keV band, it is a good candidate for
harboring an IMBH ($M\gtrsim 100 M_\odot$).  Such an option is
corroborated by the presence of a very soft X-ray spectral component
($T\sim 200$ eV) which points to a compact object of mass definitely
larger than those of Galactic Black Hole candidates.  Moreover, the
object exhibits X-ray variability on a timescale of months.

On the basis of a 19 ks {\em Chandra} exposure and accurate astrometry
of field objects, \cite{zamp04} (Z04 hereafter) derived the source
position with an uncertainty of 0.7$\arcsec$ (RA=03:18:22.34,
DEC=--66:36:03.7; $1 \sigma$ confidence level). Inside the {\em
Chandra} error box a faint optical candidate was found on a R band
image taken with the ESO 3.6 m telescope in January 2002.

Here we present a follow-up study of the optical counterpart of NGC
1313 X-2, based on photometric archive data obtained with the ESO VLT
telescope and a reconsideration of our previous photometry of the ESO
3.6 m+EFOSC image.

\section{Optical observations: data reduction and analysis}

\begin{center}
\begin{figure*}
\plottwo{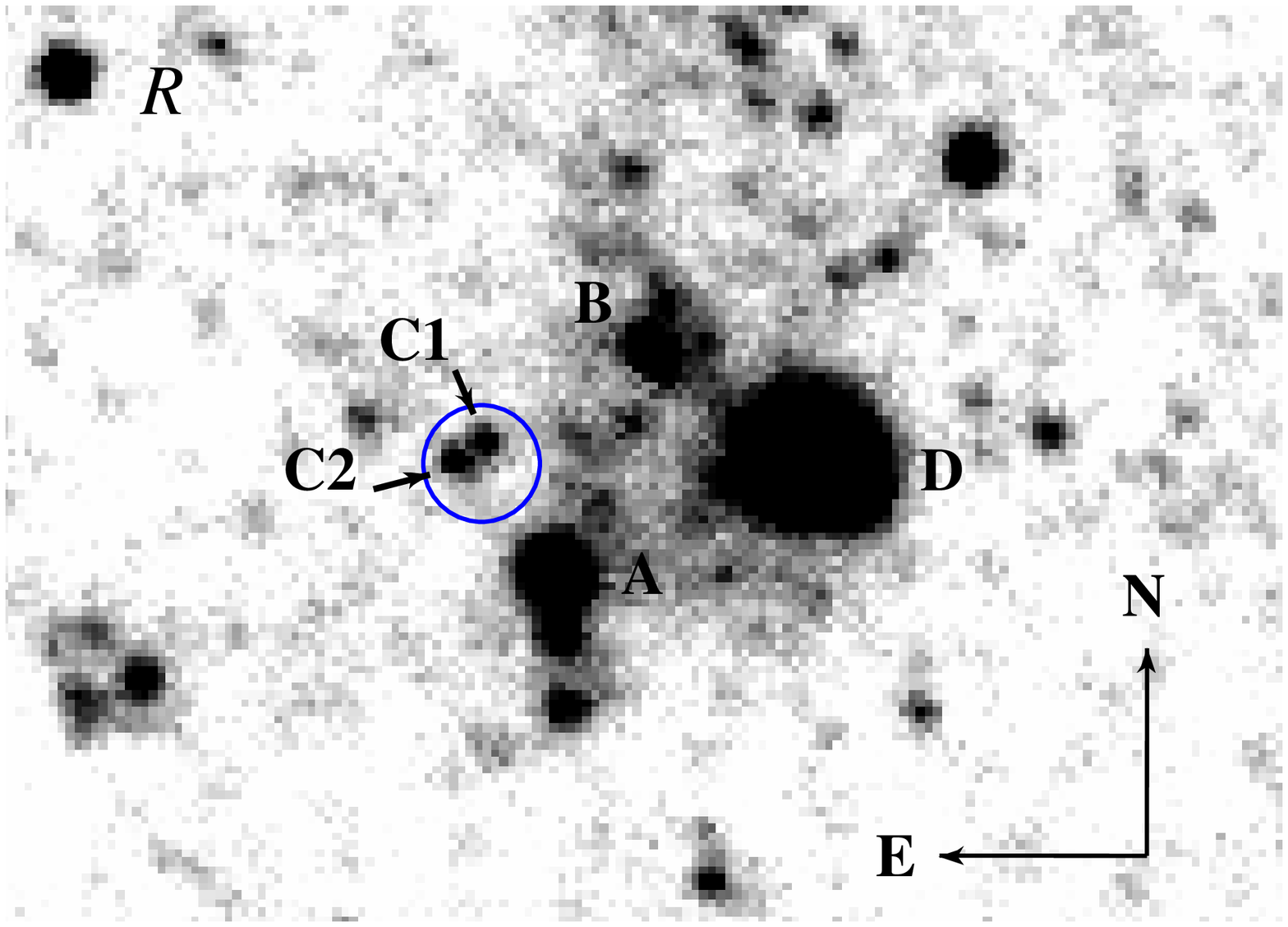}{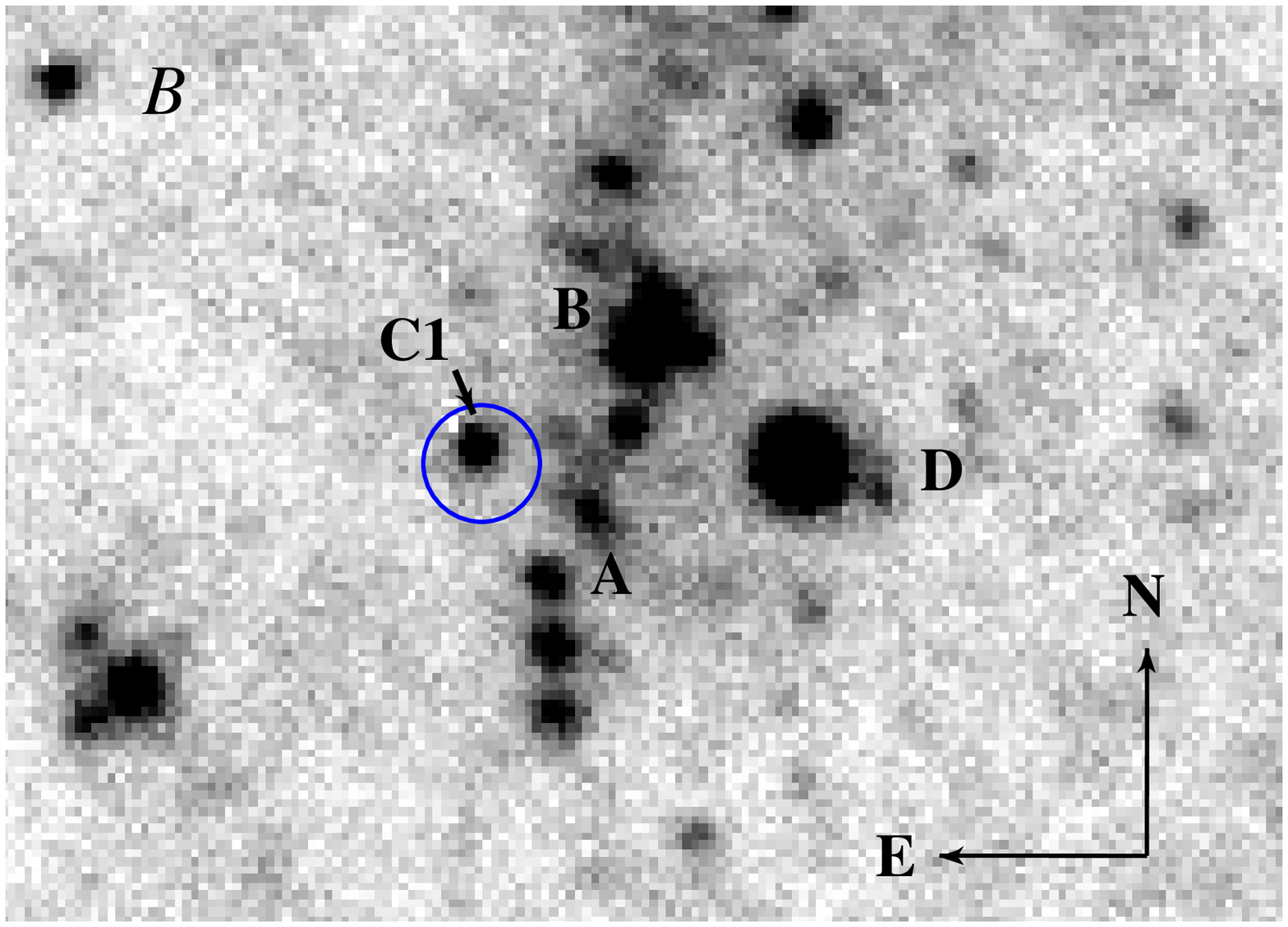}
\caption{$R$ ({\it left}) and $B$ ({\it right}) VLT+FORS1 images of the field
around NGC 1313 X-2 (30$'' \times$20$''$). The circle is the 2$\sigma$ {\it Chandra}
error-box (1.4$''$). In the $R$ frame, the counterpart is clearly resolved in
two point sources, C1 and C2.
\label{fig:BVR}}
\end{figure*}
\end{center}

We analyzed archive ESO VLT+FORS1 images ($BVR$) and spectra of NGC
1313 X-2 taken between December 2003 and January 2004 (Program ID
072.D0614). The observations are listed in Table
\ref{tab:obslog}. Data reduction was performed using standard
IRAF\footnote{IRAF is distributed by the National Optical Astronomy
Observatories, which are operated by the Association of Universities
for Research in Astronomy, Inc., under cooperative agreement with the
National Science Foundation.} procedures. The images were
astrometrically calibrated using 29 GSC2 stars. The calibration
uncertainty, tested with GSC2 stars not used for the calibration, is
$\sim 0.3 \arcsec$. 
Finally, for each band the images were combined and cleaned from
cosmic rays. Figure \ref{fig:BVR} shows the combined $R$ and $B$
images. On the same night a short exposure of the standard PG 0231+051
Landolt field \citep{land02} was also taken in each band. Aperture
photometry (1$''$ radius) was performed on the combined images. The
instrumental magnitudes were then calibrated with the Landolt standard
stars in the Bessel-Cousins system \citep[see][for extinction coefficients and
color terms]{pat03}.

Z04 give the {\it R} magnitude of a number of objects around NGC 1313 X-2
and of the proposed counterpart (object C in their paper). The latter was
close to the limit of detectability on their image and appeared as a
single object. Thanks to the higher resolution of the VLT image, in
the $R$ and $V$ exposures we are able to resolve object C in two
distinct point sources, C1 and C2. Both are inside the {\em Chandra}
error box (see Figure \ref{fig:BVR}). Object C2 is not detected in the
$B$ band frame. Magnitudes, colors and astrometric positions of the two 
candidate
counterparts, C1 and C2, and of objects A, B, and D (following Z04)
are reported in Table \ref{tab:mag}. The photometric errors are the
$2\sigma$ statistical errors on the measurements with the different
Landolt standards. For object C2, we quote an upper limit to the $B$
band magnitude using the plate limit ($B = 25.2$). 

The total magnitude of object C1+C2 is $ R=23.2\pm0.05 $.
We note that this measurement is not in agreement with the 
estimated magnitude of object C obtained in our previous  
ESO 3.6m image ($R=21.6$; see Z04).
In fact, re-analyzing the old image we found that the stellar 
background used for subtraction was underestimated, leading 
to a measurement $\sim 0.3$ mag brigther.
We also found a mistake in the adopted exposure time of the
image that causes an overestimate of another $\sim$ 0.9 mag. 
The revised magnitude of object C in the ESO 3.6m image is now found 
to be $R=22.9\pm 0.2$, in reasonable agreement (within 2$\sigma$) 
with the VLT measurement.

\begin{center}
\begin{figure}
\plotone{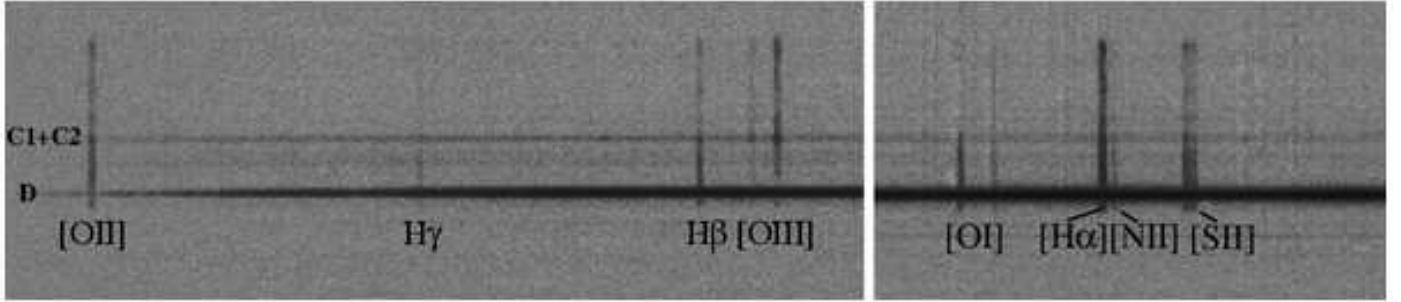}
\caption{Two-dimensional spectrum (VLT+FORS1) of objects C1+C2.
The slit (1$''$) is oriented to include object D. The wavelength
intervals are 3600-5200 A ({\it left}) and  6200-7100 A ({\it right}). 
\label{fig:neb}}
\end{figure}
\end{center}

\begin{center}
\begin{figure}
\plotone{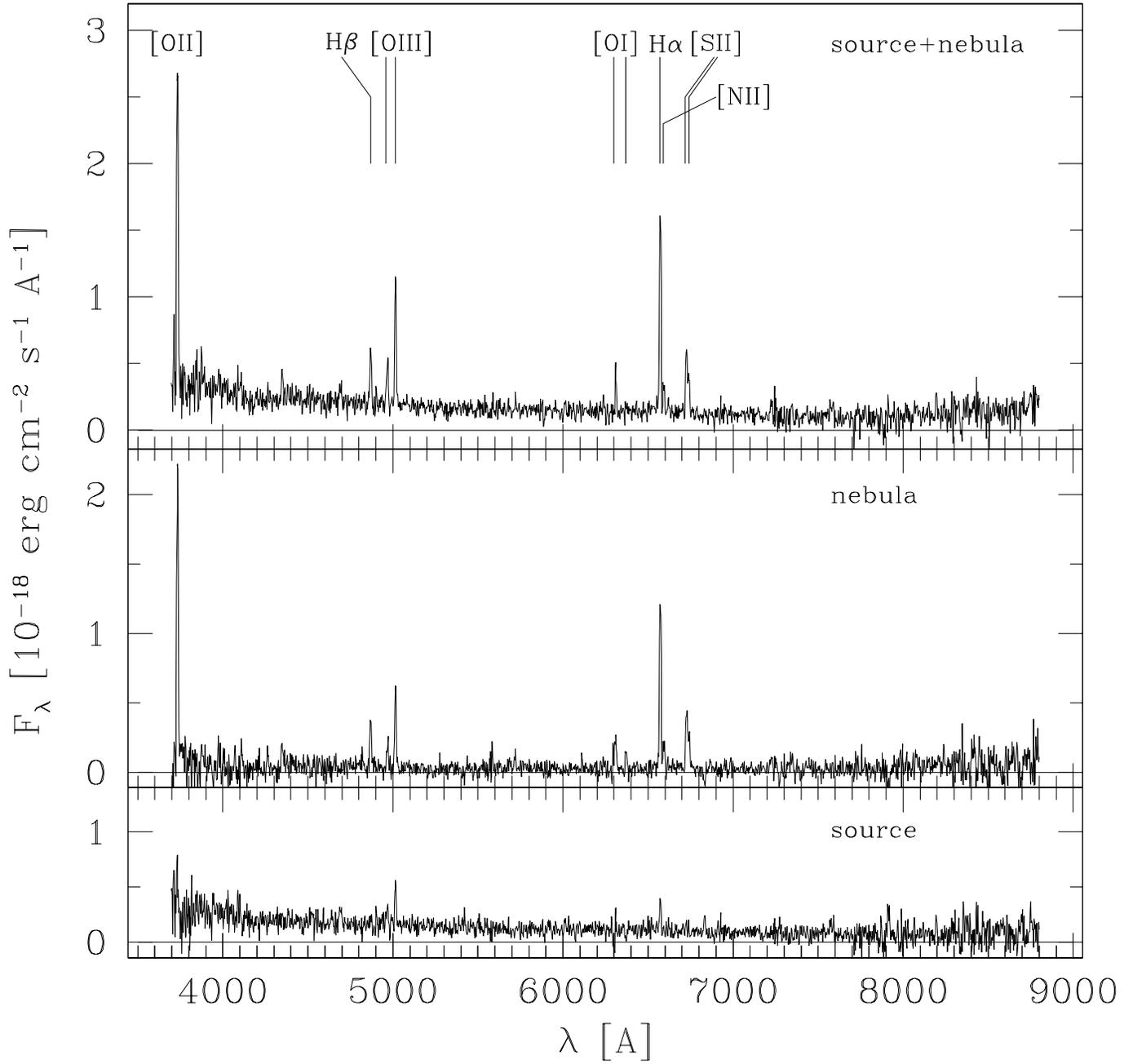}
\caption{From top to bottom: VLT+FORS1 spectrum of 15 January 2004 (3700-8900 A), extracted from an
aperture of 2.2$''$ centered on the unresolved object C1+C2;   
average spectrum of the nebula, extracted from two 1$''$ apertures located
eastward and westward of the source position and adjacent to it; 
the nebula-subtracted spectrum of object C1+C2.\label{fig:spec1d}}
\end{figure}
\end{center}

In addition to the images, we also analyzed four VLT+FORS1 spectra
($\lambda_{c}$=5900 A, $\lambda/\Delta \lambda$=440 at $\lambda_{c}$) 
of objects C1+C2 taken in different nights (Table \ref{tab:obslog}).
The slit ($1\arcsec$) was oriented to include object D. After performing 
standard 
reduction, all spectra were sky subtracted, wavelenght calibrated through 
comparison lamp exposures and flux calibrated using standard star spectra 
obtained in the same night. In these VLT+FORS1 spectra the two sources 
(C1 and C2) are not spatially resolved.  The 2D spectrum taken on 15 
January 2004 is shown in Figure \ref{fig:neb}. Nebular emission lines 
of [OII] $\lambda$ 3727 A, H$_\gamma$, H$_\beta$, [OIII] $\lambda\lambda$ 
4959-5007 A, [OI] $\lambda$ 6300 and 6364 A, H$_\alpha$, [NII] $\lambda$ 
6583 A and [SII] $\lambda\lambda$ 6717-6731 A are clearly detected. 
Note that this is the first detection of a [OII] line from this nebula.
A one dimensional spectrum was extracted over an aperture of 2.2$''$
centered on object C1+C2 from each of the four combined spectra. Two 
nebular spectra were extracted from different 1$''$ apertures, eastward 
and westward of the source position and adjacent to it. The two spectra 
were then averaged and the resulting spectrum subtracted from that of 
object C1+C2.  
All these spectra, taken on January 15th, are shown in Figure \ref{fig:spec1d}.  
The nebula-subtracted source spectrum shows no evident emission or absorption 
lines. Residuals are present in coincidence with some nebular lines 
(expecially [OIII] and H$_\alpha$), with an upper limit to the equivalent 
width of $\sim 30$ A.
In particular the residual flux in the [OIII] line is a non negligible fraction
of the nebular flux. This appears to be caused by an increased emission of
the nebular line around the position of object C1+C2. It is not clear if this is
simply induced by a change in the rather irregular spatial profile of the
nebular line or by a variation of the physical conditions produced by the
presence of the nearby ULX. Finally, marginal evidence of an excess in 
emission may be seen at 4686 A, corresponding to HeII emission, but the line is
not statistically significant.

\section{Results}\label{disc}

The superb quality of the VLT images reveals that two distinct
objects, C1 and C2, are visible inside the {\em Chandra} error box of
NGC 1313 X-2 in the {\it R} and {\it V} bands. From the astrometric positions reported 
in Table \ref{tab:mag}, we infer a separation of 
$0.75\arcsec$ and a position angle (C2 with respect to C1) of
$\sim$131$^\circ$ . Note that the two objects were unresolved in our 
previous 3.6 m+EFOSC image. 

The possibility that either C1 or C2 may be a background AGN appears
very unlikely. In fact, no statistically significant emission line at
wavelengths longer than H$_\alpha$ is observed in the optical spectrum
nor any other feature that may be identified with a highly redshifted
emission line.
Furthermore, from the {\it R} magnitude of objects C1 and C2, and the X-ray
flux of the last {\em XMM-Newton} observation ($f_X \sim 2\times
10^{-12}$ erg cm$^{-2}$ s$^{-1}$; Z04), we obtain a X-ray to optical
flux ratio $f_X/f_R \ga 2000$.  This is more than an order of
magnitude larger than the typical value for an AGN ($\lesssim
100$).
The possibility that object C2 is a highly obscured AGN is not very
convincing either. In fact, to produce the same X-ray flux but a
$f_X/f_R$ ratio of $\sim 100$, the $R$ magnitude should be $\sim
20$. Therefore the additional reddening required to make object C2
look like an obscured AGN is $A_R
\simeq 3.5$ mag, corresponding to a column density $\sim 9\times
10^{21}$ cm$^{-2}$ \citep{bol78}, much larger than that inferred from
X-ray spectral fits [$\sim (3_{-0.4}^{+0.9})\times {10^{21}}$
cm$^{-2}$]. Unless absorption in NGC 1313 is characterized by an
extinction law quite different with respect to that in our Galaxy, we
are led to rule out an identification with an (obscured) AGN.

\begin{center}
\begin{figure}
\plotone{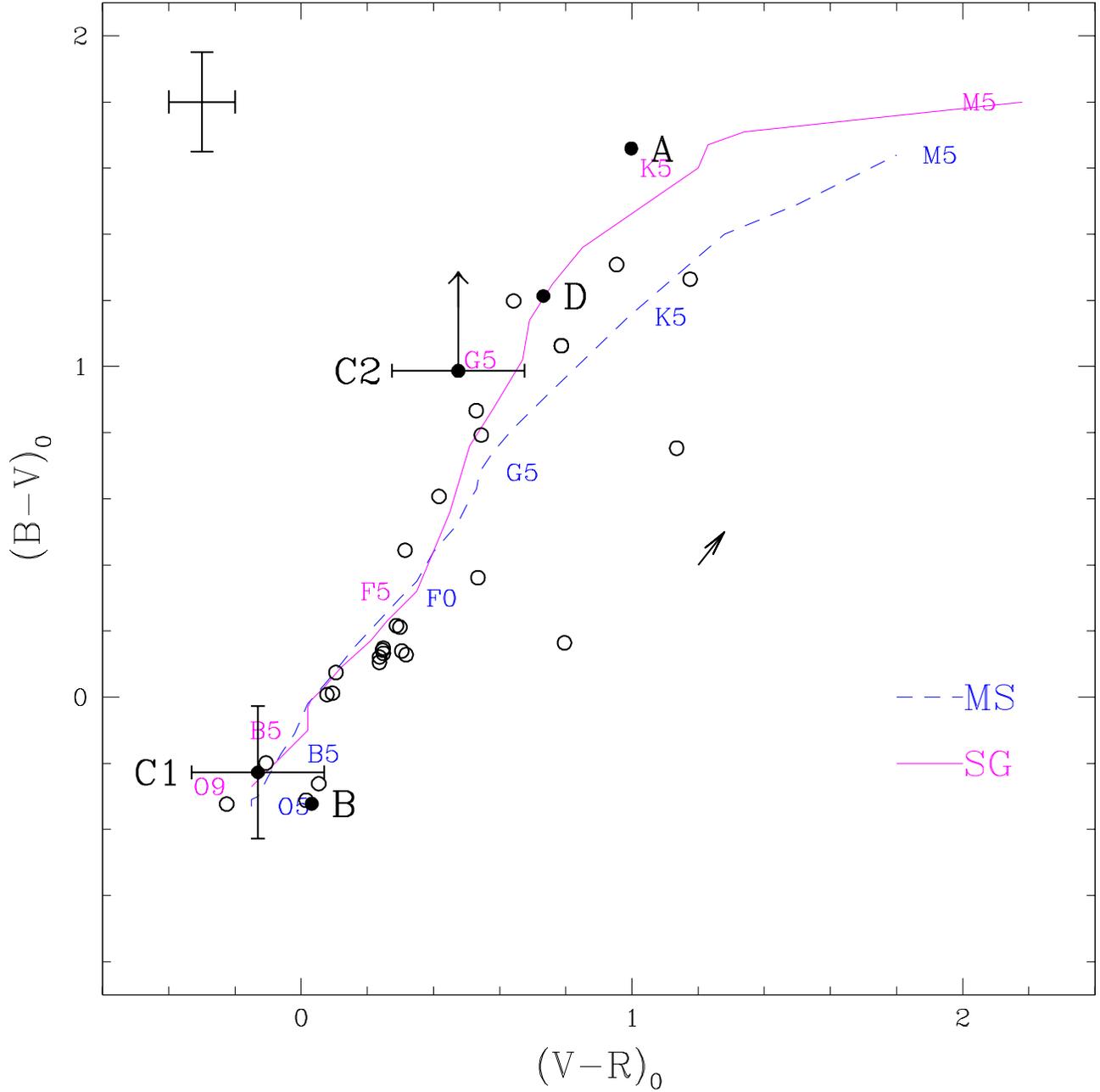}
\caption{$(B-V)_0$ vs. $(V-R)_0$ for a sample of field objects around NGC 1313 X-2,
including A, B, C1, C2 and D. Error bars are shown at the upper left corner. 
Measurements were corrected for
Galactic exctinction ($A_V=0.3$, $E_{B-V}=0.1$, corresponding to a
column density $N_H=6\times 10^{20}$ cm$^{-2}$). The arrow indicates the
reddening vector corresponding to $E_{B-V}=0.1$. The solid and dashed lines
represent the colors of supergiant (SG) and main sequence (MS) stars.\label{fig:colcol}}
\end{figure}
\end{center}

A photometric analysis of a large sample of field stars ($> 30$) was
performed on the three $BVR$ images. A color-color plot of these
objects is shown in Figure \ref{fig:colcol}. Almost all of them have
stellar colors, apart from two on the lower right part of the diagram
that are bluer than ordinary stars. Given their apparent magnitude,
objects earlier than spectral type F (including C1) are stars in NGC
1313, while the others may be Galatic foreground stars or belong to
NGC 1313. However, object C2 cannot be a Galatic foreground star because its
absolute visual magnitude would be $M_V \sim 9$, too large to be
consistent with its colors.

Within the photometric errors, the colors of object C1 appear to be
consistent with those of a A3-O9 I or a A2-B0 V star, while those of
C2 with a G8-G7 I star (see e.g. \citealt{all00,brad82}). Unfortunately, the 
optical continuum does not
provide useful information for assessing the spectral type because the
light from both objects contributes to it. Observationally, the slope of the
continuum can be caracterized by a power law, $\lambda^{-1.8}$. 
The absence or extreme weakness of the HeII $\lambda$ 4686 A emission line 
in the optical spectrum  suggests that X-ray irradiation is not dominant. 
Taking Galactic absoption
into account and assuming $A_V \simeq 0.3$ (\citealt{card89} exctinction law
with $R_V=A_V/E_{B-V}=3.1$ has been used throughout), the unreddened
colors of object C1 are $(V-R)_0=-0.2 \pm 0.2$ and $(B-V)_0=-0.2 \pm
0.2$, consistent with those of a B8-O I or A0-O5 V star (see Figure
\ref{fig:colcol}). For object C2 it is $(V-R)_0=0.4 \pm 0.2$ and
$(B-V)_0 \ga 1.0$, consistent only with a G4 I star 
(see again Figure \ref{fig:colcol}). Recently, \cite{liu05b} performed
a 6.4 m Magellan/Baade observation of the field around NGC 1313 X-2
and found a $I=23.3$ mag object in coincidence with the position of
C1+C2 (that appear unresolved in their $I$ frame). Assuming that the
flux in the $I$ band originates mainly from the redder object C2, we
then obtain $(R-I)=0.3\pm 0.2$, consistent with our tentative spectral
classification.

At the distance of NGC 1313 ($d=3.7$ Mpc; \citealt{tul88}) the {\it V}
magnitudes of C1 and C2 (reported in Table~\ref{tab:mag}) translate
into the absolute magnitudes $M_V\sim -4.6$ and $\sim -4.1$,
respectively. Comparing these values with the absolute magnitudes of
main sequence and supergiant stars (e.g. \citealt{all00,brad82}), we find
that the observed value is consistent only with a B0-09 main sequence
star for C1, while it is consistent with a G4 supergiant of type Ib for C2.
Therefore, we
conclude that both C1 and C2 are stars in NGC 1313, with C1 an early type
main sequence star and C2 a supergiant. The bolometric
luminosities of the two objects are $\sim 3\times 10^{38}$ erg s$^{-1}$
and $\sim 2\times 10^{37}$ erg s$^{-1}$, respectively.

Given the density of objects in the field of view ($\sim 50-100$
arcmin$^{-2}$), a significant fraction of which are supergiants in NGC
1313, the probability that C1 or C2 fall by chance inside the
2$\sigma$ {\it Chandra} error box is not negligible ($\sim
0.1$). However, the chance occurrence of two objects, separated by
only 0.7$''$, inside the X-ray error box is $\sim 5 \times 10^{-3}$,
sufficiently small to be considered rather unlikely. Actually, if both
C1 and C2 are stars in NGC 1313, a physical association may not
be unplausible (the distance corresponding to the apparent separation
on the sky is $\sim 10$ pc). Therefore, we conclude that the ULX is
most probably physically associated to either object C1 or
C2. Irrespectively of which of the two objects is the actual
counterpart, NGC 1313 X-2 appears to be a high mass X-ray binary with
a very massive donor star.

We now briefly consider what are the consequences if either C1 or C2
are the counterparts of the X-ray source. A B0-O9 main sequence star
has an initial mass of $\sim 20 M_\odot$. In this respect our analysis
essentially confirms the original suggestion by Z04, who proposed that
the optical counterpart of NGC 1313 X-2 may be an O type main sequence
star in NGC 1313.
If the colors are affected by the binary interaction, the estimated
mass may vary somewhat.
A $\sim 20 M_\odot$ donor star could easily provide the mass transfer
rate required to fuel the accreting black hole through Roche-lobe
overflow during the main sequence phase if the orbital separation is
$\approx$ 1 AU.  In these conditions, the mass transfer would be
stable and the source persistent (\citealt{patr05}; Patruno \& Zampieri, in
preparation).

If object C2 is the counterpart, the nature of the system remains
unchanged. In this case X-ray irradiation may be significant and give a
non-negligible contribution to the optical emission. 
The mass corresponding to a G4
 supergiant is $\sim 10 M_\odot$. The same caveat discussed for C1
about the possibility that the colors and mass estimate are affected
by binary interaction applies also in this case. The mass transfer rate
provided by such a donor star throug Roche-lobe overflow is certainly
adequate also for large orbital separations. Wind accretion may also
be a viable alternative.

At the present stage, there seems to be no compelling argument to rule
out any of the two objects C1 and C2 as possible counterparts of NGC
1313 X-2. 
In both cases, NGC
1313 X-2 is a high-mass X-ray binary with a very massive donor. In
order to distinguish which of the two objects is the counterpart,
further observations are needed. In particular, we aim at optical
spectroscopy under optimal seeing conditions and with the slit along
the line connecting C1 and C2, which would allow us to check the existence 
of typical lines of X-ray binaries (like He II $\lambda$ 4686 A). 
Theoretical modelling
(binary evolution calculations) may also help distinguish the
counterpart. It will also clarify the properties of the accretion flow
and the compatibility with the IMBH or stellar black hole
interpretations.

\acknowledgements
Based on observations made with ESO Telescopes at the Paranal 
Observatories, obtained from the ESO/ST-ECF Science Archive Facility.
This work has been partially supported by the Italian Ministry for
Education, University and Research (MIUR) under grant PRIN-2004-023189.
We thank Alessandro Patruno for interesting discussions on the properties of
IMBH binary sistems.


\begin{table}
\begin{center}
\caption{Log of the VLT+FORS1 observations.
\label{tab:obslog}}
\begin{tabular}{ccccc}
\tableline\tableline
Obs. type&Date & Filter/Grism($\lambda_{c}$)& Exp. time (s)& Seeing ($\arcsec$)\\
\tableline
Image&2003-12-24&B	  &840$\times$2  &1.0\\
Image&2003-12-25&V	  &600$\times$2  &0.8\\
Image&2003-12-24&R	  &500$\times$2  &0.8\\
\tableline
Spec.&2003-12-22&5900 A&1300$\times$2 &...\\
Spec.&2003-12-24&5900 A&1300$\times$2 &...\\
Spec.&2003-12-30&5900 A&1300$\times$2 &...\\
Spec.&2004-01-15&5900 A&1300$\times$2 &...\\
\tableline
\end{tabular}
\end{center}
\end{table}


\begin{table}
\begin{center}
\caption{Astrometric positions, magnitudes and colors of the sources around NGC 1313 X-2 (see Figure \ref{fig:BVR}).
\label{tab:mag}}
\footnotesize
\begin{tabular}{cccccccc}
\tableline\tableline
Source & RA & DEC & B & V & R & B-V & V-R \\
\tableline
A &03:18:21.97$\pm$0.05&-66:36:06.4$\pm$0.3&23.5$\pm$0.15 &21.7$\pm$0.05 &20.6$\pm$0.05 & 1.8$\pm$0.15   & 1.1$\pm$0.1 \\
B &03:18:21.57$\pm$0.05&-66:36:00.8$\pm$0.3&22.4$\pm$0.15 &22.7$\pm$0.05 &22.5$\pm$0.05 &-0.3$\pm$0.15   & 0.2$\pm$0.1 \\
C1&03:18:22.26$\pm$0.05&-66:36:03.3$\pm$0.3&23.5$\pm$0.15 &23.6$\pm$0.15 &23.7$\pm$0.15 &-0.1$\pm$0.2   &-0.1$\pm$0.2 \\
C2&03:18:22.36$\pm$0.05&-66:36:03.8$\pm$0.3&$\ga$25.2       &24.1$\pm$0.15 &23.6$\pm$0.15 & $\ga$1.1	& 0.5$\pm$0.2 \\
D &03:18:20.96$\pm$0.05&-66:36:03.6$\pm$0.3&20.3$\pm$0.15 &18.9$\pm$0.05 &18.1$\pm$0.05 & 1.4$\pm$0.15   & 0.8$\pm$0.1 \\
\tableline
\end{tabular}
\end{center}
\end{table}


\begin{thebibliography}{}
\bibitem[\protect\citeauthoryear{Braddley}{1982}]{brad82} Braddley, S., 
1982, Numerical data and functional relationships in science and 
technology, New Series, group VI, vol. 2, Astronomy and Astrophysics, Extension
 and Supplement to vol. 1, Stars and Star Clusters, Springer
\bibitem[\protect\citeauthoryear{Cox}{2000}]{all00} Cox, A.N. 2000, Allen's
Astrophysical Quantities, Springer
\bibitem[\protect\citeauthoryear{Bohlin et al.}{1978}]{bol78} Bohlin, R.C., 
Savage, B.D., \& Drake, J.F. 1978, \apj, 224, 132
\bibitem[\protect\citeauthoryear{Cardelli et al.}{1989}]{card89} Cardelli, J.A., Clayton, G.C.
\& Mathis, J.S. 1989, \apj, 345, 245
\bibitem[\protect\citeauthoryear{Colbert \& Ptak}{2002}]{col02} Colbert, E.J.M.,
\& Ptak, A.F. 2002, \apjs, 143, 25
\bibitem[\protect\citeauthoryear{Foschini et al.}{2002a}]{fosc02a} Foschini, L., et al.
2002a, \aap, 392, 817
\bibitem[\protect\citeauthoryear{Foschini et al.}{2002b}]{fosc02} Foschini, L.,
Ho, L.C. \& Masetti, N. 2002b, \aap, 396, 787
\bibitem[\protect\citeauthoryear{Kaaret et al.}{2003}]{kaa03}
Kaaret, P., Corbel, S., Prestwich, A.H., \& Zezas, A. 2003, Science, 299, 365
\bibitem[\protect\citeauthoryear{Kaaret et al.}{2004}]{kaa04}
Kaaret, P., Ward, M.J., \& Zezas, A. 2004, \mnras, 351, 83
\bibitem[\protect\citeauthoryear{King et al.}{2001}]{kin01} King, A.R. et al., 
2001, \apjl, 552, 109
\bibitem[\protect\citeauthoryear{Landolt}{1992}]{land02} Landolt, A.U. 1992,
\aj, 104, 340
\bibitem[\protect\citeauthoryear{Liu et al}{2002}]{liu02} Liu, J., Bregman,
J.N., \& Seitzer, P. 2002, \apjl, 580, 31
\bibitem[\protect\citeauthoryear{Liu et al}{2004}]{liu04} Liu, J., Bregman,
J.N., \& Seitzer, P. 2004, \apj, 602, 249
\bibitem[\protect\citeauthoryear{Liu \& Bregman}{2005a}]{liu05} Liu, J., \& Bregman,
J.N.  2005, \apjs, 157, 59
\bibitem[\protect\citeauthoryear{Liu et al}{2005b}]{liu05b} Liu, J., Bregman,
J.N., Seitzer, P. \& Irwin, J. 2005, astro-ph/0501310
\bibitem[\protect\citeauthoryear{Masetti et al.}{2003}]{mas03}
Masetti, N., et al. 2003, \aap, 406, L27
\bibitem[\protect\citeauthoryear{Miller et al.}{2003}]{mil03} Miller, J.M.,
Fabbiano, G., Miller, M.C., \& Fabian, A.C. 2003, \apjl, 585, 37
\bibitem[\protect\citeauthoryear{Miller et al.}{2005}]{miln05} Miller, N.A.,
Mushotzky, R.F. \& Neff, S.G. 2005, \apjl, 623, 109
\bibitem[\protect\citeauthoryear{Pakull \& Mirioni}{2002}]{pak02} Pakull, M.W.,
\& Mirioni, L. 2002, in Proc. ESA Symp., New Visions of the X-ray
Universe in the {\it XMM-Newton\/} and {\it Chandra\/} Era, eds.
F. Jansen et al. (ESA SP-488) (astro-ph/0202488)
\bibitem[\protect\citeauthoryear{Patat}{2003}]{pat03} Patat, F. 2003, \aap, 400,
1183
\bibitem[\protect\citeauthoryear{Patruno et al.}{2005}]{patr05}  Patruno, A., 
Colpi, M., Faulkner, A. \& Possenti, A., 2005, \mnras ~in press (astro-ph/0507229)
\bibitem[\protect\citeauthoryear{Swartz et al.}{2004}]{swa04} Swartz, D.A.,
Ghosh, K.K., Tennant, A.F., \& Wu, K. 2004, \apjs, 154, 519
\bibitem[\protect\citeauthoryear{Tully}{1988}]{tul88} Tully, R.B. 1988, Nearby
Galaxies Catalog (Cambridge: Cambridge University Press)
\bibitem[\protect\citeauthoryear{Turolla et al.}{2005}]{tur05} 
Turolla, R. et al. 2005, in Proceedings of the 35th COSPAR Scientific
Assembly, p. 3749
\bibitem[\protect\citeauthoryear{Zampieri et al.}{2004}]{zamp04}
Zampieri, L. et al. 2004, \apj, 603, 523 (Z04)
\end{thebibliography}
\end{document}